\documentclass[10pt,a4paper,5p, oneside, preprint]{elsarticle}

\usepackage[english]{babel}
\usepackage{blindtext}
\usepackage{amsmath,amssymb,amsthm,amstext,amsfonts}
\usepackage{color}
\usepackage{units}
\usepackage[latin1]{inputenc}
\usepackage{acronym}
\usepackage{subfigure}
\usepackage{booktabs,longtable}
\usepackage{multirow}
\usepackage{enumitem}
\usepackage{microtype}
\usepackage[switch]{lineno}
\usepackage[pdftex,plainpages=false,pdfpagelabels]{hyperref}

\begin{document}

\title{Towards Searching for Photons with Energies beyond\\the PeV Range from Galactic PeVatrons}
\author[si]{M. Niechciol\corref{cor1}}
\ead{niechciol@physik.uni-siegen.de}
\author[si]{C. Papior}
\author[si]{M. Risse}

\address[si]{Center for Particle Physics Siegen, Department of Physics, University of Siegen, Germany}

\cortext[cor1]{Corresponding author}

\begin{abstract}

Several gamma-ray observatories have discovered photons of cosmic origin with energies in the PeV ($\unit[10^{15}]{eV}$) range. Photons at these energies might be produced as by-products from particle acceleration in so-called PeVatrons, which are widely assumed to be the sources of a large part of galactic cosmic rays. Based on recent measurements of these PeV $\gamma$-sources by LHAASO and HAWC, we extrapolate the energy spectra of selected sources up to the ultra-high-energy (UHE, $\geq$$\unit[10]{PeV}$) regime. The goal of this study is to evaluate if (and under what conditions) giant air-shower observatories, for example the Pierre Auger Observatory, could contribute to testing the UHE luminosity of PeV $\gamma$-sources. Possible propagation effects are investigated as well as the required discrimination power to distinguish photon- and hadron-initiated air showers. For present detector setups, it turns out to be challenging to achieve the required sensitivity due to the energy threshold being too high or the detection area too small. Dedicated detector concepts appear to be needed to explore the UHE frontier. Ultimately, this could provide complementary information on the sources of cosmic rays beyond the PeV regime---a key objective of current efforts in multimessenger astronomy.
\end{abstract}


\maketitle


\section{Introduction}
\label{sec:introduction}

In the past years, cosmic photons of ever-increasing energy have been observed. For some time, the maximum photon energies were in the range of $\unit[0.01-0.1]{PeV}$ ($\unit[1]{PeV} = \unit[10^{15}]{eV}$), measured by imaging atmospheric Cherenkov telescopes like H.E.S.S.~\cite{HESS:2006fka}, MAGIC~\cite{MAGIC:2011kvf}, or VERITAS~\cite{Weekes:2001pd}. The advent of wide-field gamma-ray observatories of large duty cycle like Tibet AS$\gamma$~\cite{TibetASg:2019ivi}, HAWC~\cite{HAWC:2023opo}, and LHAASO~\cite{Cao:2010zz} paved the way for pushing the energy frontier. LHAASO and HAWC reported the observation of photons with energies in the PeV range from a number of (galactic) sources~\cite{LHAASO:2021gok,HAWC:2019tcx}. These ``PeV $\gamma$-sources'' are of considerable interest in the context of multimessenger astronomy, in particular for identifying the sources of cosmic rays in the PeV range and above, which is a key objective of current efforts in this rapidly emerging field.

The sites where the acceleration of cosmic rays to at least PeV energies could take place are commonly termed ``PeVatrons''. Although the exact mechanisms responsible for the acceleration of the initial particles in these PeVatrons are not yet fully understood, it is expected that photons are emitted as by-products of the acceleration through different processes, either involving leptons or hadrons~(see, e.g.,~\cite{Casanova:2022wva,Cristofari:2021jkl,Cardillo:2023hbb}). Leptonic processes include Bremsstrahlung and inverse Compton scattering, while hadronic processes include interactions of the initial particles with matter or surrounding radiation fields, producing neutral pions which then decay into photons. 
The photons reach energies of about a factor 3$-$10 below those of the charged particles that are accelerated in such PeVatrons. Hence, photons observed from PeV $\gamma$-sources are indicative of charged-particle acceleration to energies beyond the PeV scale at these sites.
These processes can occur, for example, at supernova remnants, pulsars, and pulsar wind nebulae (such as the Crab), making them possible PeVatrons---in fact, they are widely assumed to be the sources of a large part of galactic cosmic rays---but microquasars, superbubbles, or young massive starclusters are among the PeVatron candidates as well (see, e.g.,~\cite{Casanova:2022wva,Cristofari:2021jkl,Cardillo:2023hbb}).

Some of the observed PeV $\gamma$-sources exhibit photon spectra without a cutoff up to the maximum energies reached by current measurements, prompting the question up to which energies these spectra extend. Are photon energies of, e.g., $\unit[10]{PeV}$ or more reached by some sources or is there a flux suppression at a few PeV? 
In fact, due to features in the cosmic-ray spectrum and the distribution of arrival directions, the transition between galactic and extragalactic cosmic rays is assumed to occur between $\unit[100]{PeV}$ and a few $\unit[1000]{PeV}$~\cite{Coleman:2022abf}. How close, in terms of energy, are PeV $\gamma$-sources approaching this transition region? Learning about the presence or absence of photons beyond the PeV range from galactic PeV $\gamma$-sources would be of enormous interest. Specifically, detecting such photons could be a breakthrough for understanding the high-energy end of the galactic cosmic-ray production.

Thus, from the experimental side the question arises of what the prospects are for searching for photons beyond the PeV range with current and future instruments, and for possibly pushing the energy frontier by another order of magnitude to the ultra-high energy (UHE, $E \geq \unit[10]{PeV}$ in the context of this paper) regime. For HAWC and LHAASO, the exposure for UHE searches is limited due to their size~($\mathcal{O}(\unit{km^2})$). Giant air-shower experiments such as the Pierre Auger Observatory~\cite{PierreAuger:2015eyc}, mainly targeting the highest-energy cosmic rays, have recently begun to lower the energy threshold for photon searches to below $\unit[100]{PeV}$~\cite{PierreAuger:2022gkb,PierreAuger:2023nkh}. In this paper, we investigate if, and under which conditions, giant air-shower observatories could contribute to testing the UHE luminosity of PeV $\gamma$-sources. We extrapolate the measured power-law spectra of selected PeV $\gamma$-sources reported by LHAASO and HAWC to the UHE regime, and estimate the prospects of observing UHE photons from such sources or, in case of their absence, of giving valuable constraints on their flux with detectors like the Pierre Auger Observatory. 

The paper is organized as follows: In Sec.~\ref{sec:currentmeasurements}, current measurements in the PeV range published by LHAASO and HAWC are briefly summarized for four exemplary sources, including the Crab and the Galactic Center. In Sec.~\ref{sec:spectra}, we take the measured spectra for these four sources and extrapolate them beyond the PeV range. These extrapolations are used in Sec.~\ref{sec:prospects} to estimate the prospects for testing the UHE luminosity of such sources with giant air-shower observatories. The paper closes with a discussion of the results and a brief outlook (Sec.~\ref{sec:summary}).


\section{Current Measurements in the PeV Range}
\label{sec:currentmeasurements}

\begin{figure*}[p]
  \centering
  \subfigure[]{\label{fig:measurements_1}\includegraphics[width=0.49\textwidth,]{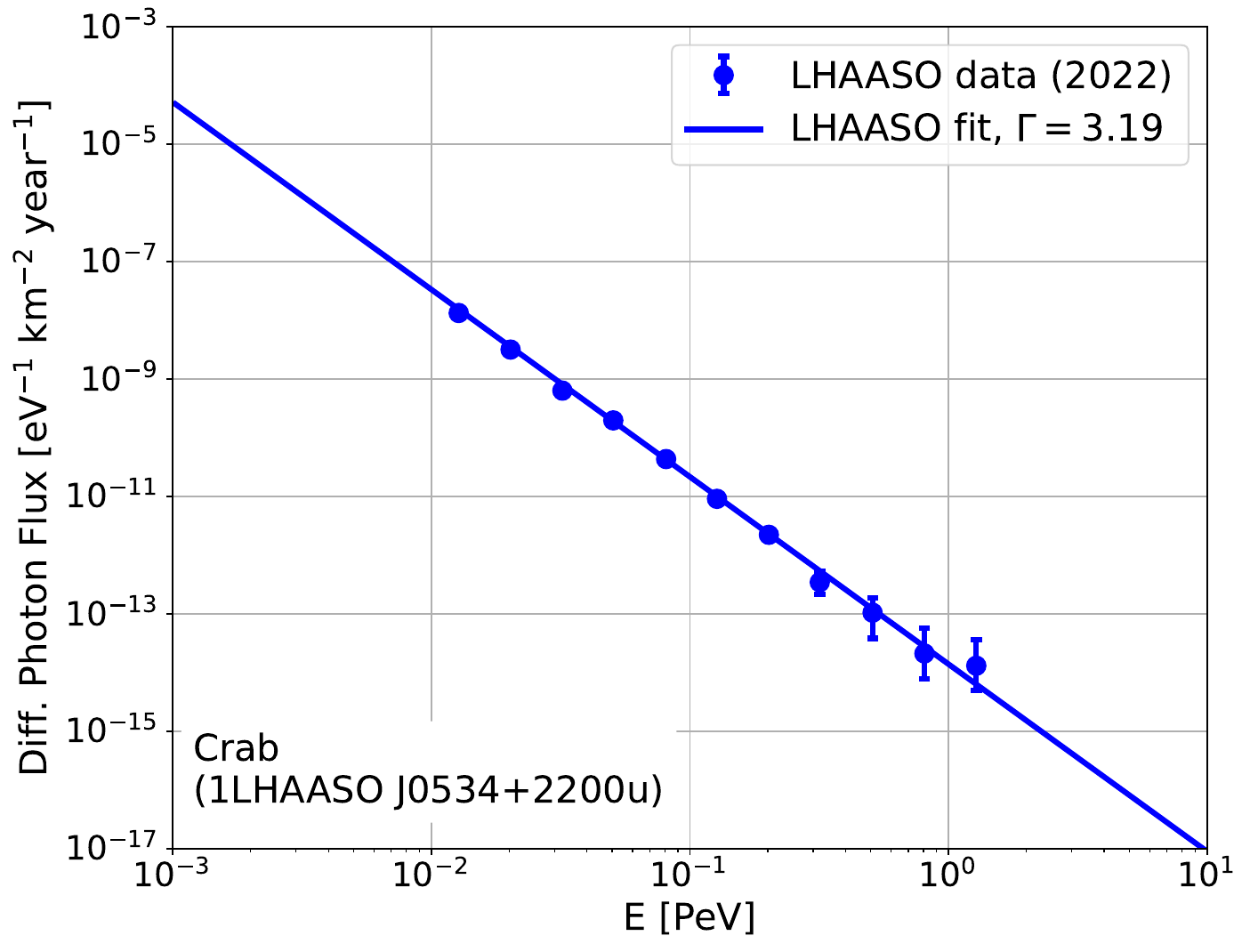}}
  \subfigure[]{\label{fig:measurements_4}\includegraphics[width=0.49\textwidth,]{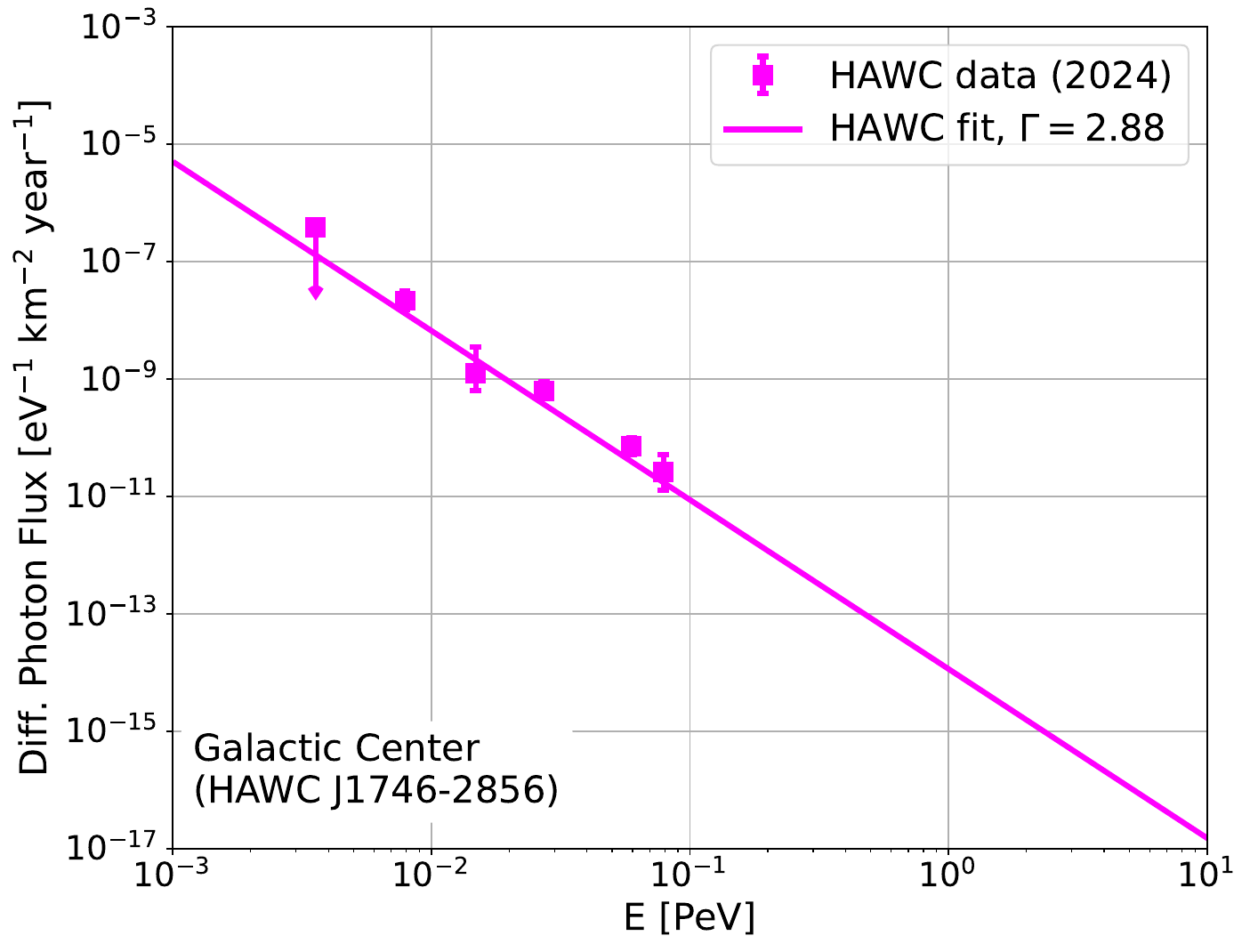}}\\
  \subfigure[]{\label{fig:measurements_2}\includegraphics[width=0.49\textwidth,]{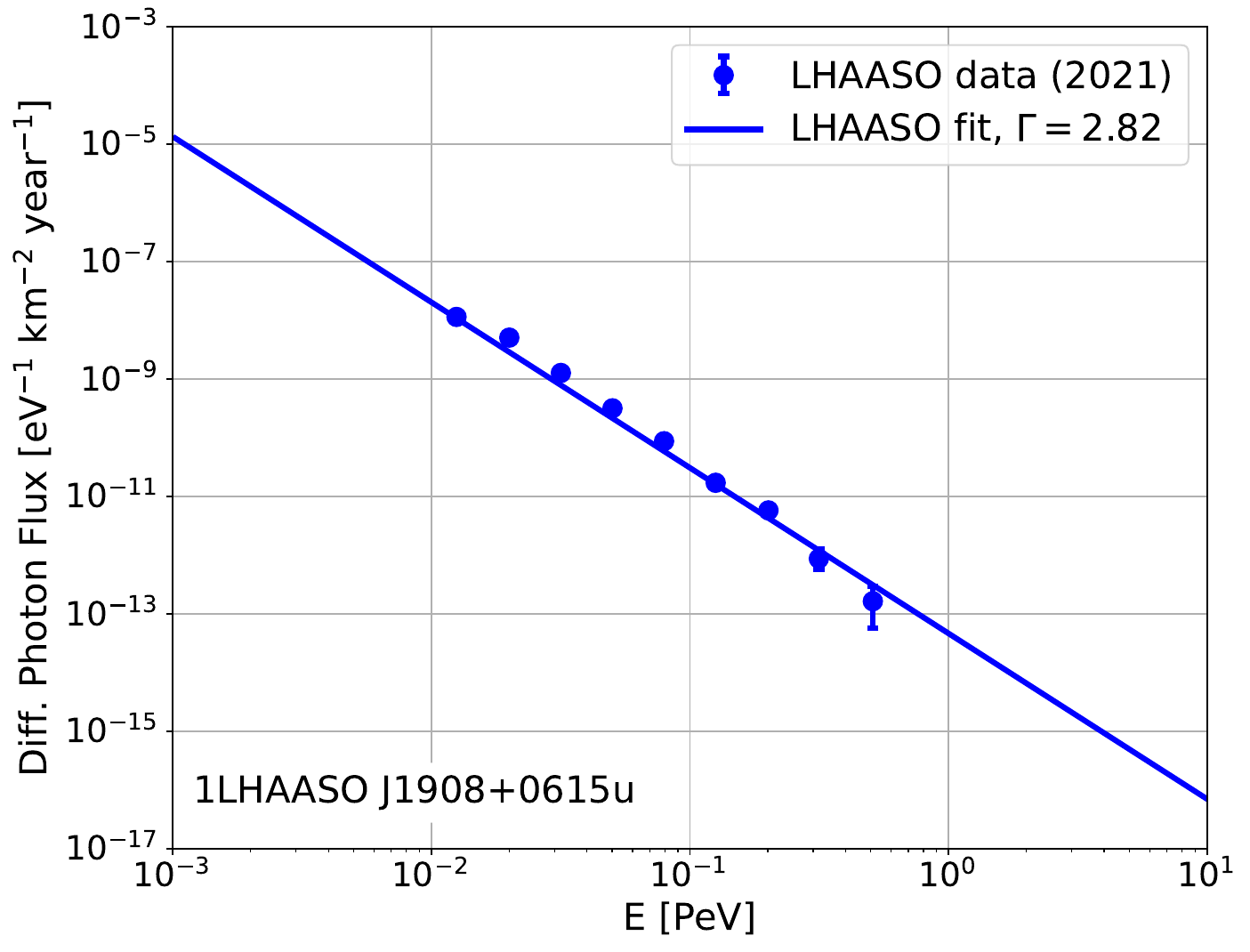}}
  \subfigure[]{\label{fig:measurements_3}\includegraphics[width=0.49\textwidth,]{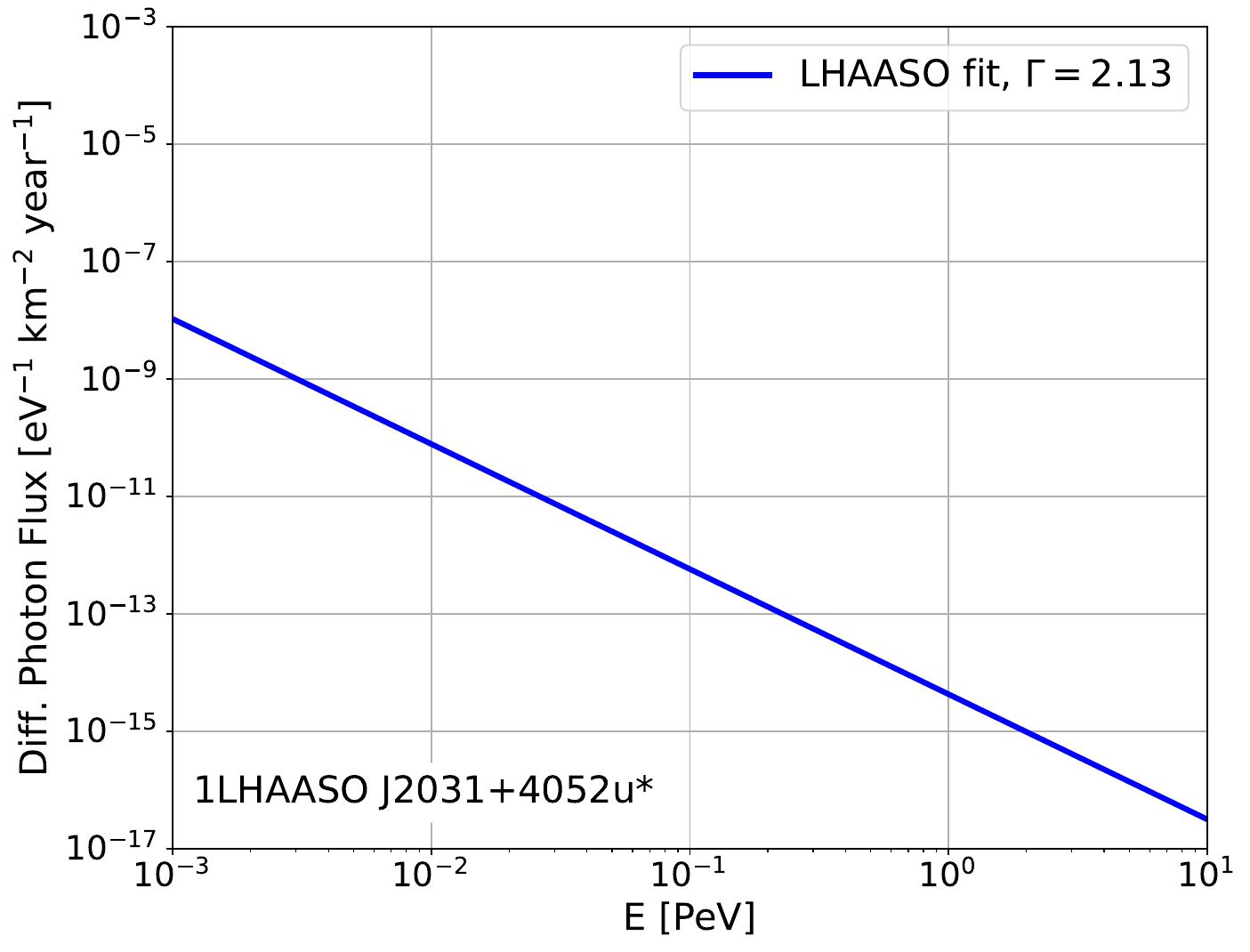}}
\caption{Energy spectra of three exemplary sources from the LHAASO catalog of galactic sources of photons with energies in the PeV range as well as the Galactic Center: \subref{fig:measurements_1}~Crab (1LHAASO J0534+2200u)~\cite{LHAASO:2023rpg,LHAASO:2021cbz}, \subref{fig:measurements_4}~Galactic Center (HAWC J1746$-$2856)~\cite{Albert:2024aaa}\subref{fig:measurements_2}~1LHAASO J1908+0615u~\cite{LHAASO:2023rpg,LHAASO:2021gok}, and \subref{fig:measurements_3}~1LHAASO J2031+4052u*~\cite{LHAASO:2023rpg}. Where available, data points are shown in addition to the power-law fits reported by the Collaborations.}
\label{fig:measurements} 
\end{figure*}

The main basis of this study are recent detections of photons in the PeV range from galactic sources, published by LHAASO and HAWC. The first LHAASO catalog of gamma-ray sources~\cite{LHAASO:2023rpg} contains in total 90 sources, 43 of which---including the Crab---show significant photon emission above $\unit[0.1]{PeV}$, expanding a previous catalog of twelve sources with photon energies above $\unit[0.1]{PeV}$ and up to $\unit[1.4]{PeV}$~\cite{LHAASO:2021gok}. These are among the highest photon energies ever measured. Additionally, the HAWC Collaboration recently reported the measurement of photons with energies exceeding $\unit[0.1]{PeV}$ from the Galactic Center~\cite{Albert:2024aaa}. In Fig.~\ref{fig:measurements}, we show exemplarily the spectra of three of the LHAASO sources---the Crab (1LHAASO J0534+2202), 1LHAASO J1908+0615u, and 1LHAASO J2031+4052u*---as well as of the Galactic Center (HAWC J1746$-$2856). 

These sources have been selected for different reasons. The Crab is an extraordinary source, playing a central role in astrophysics (for a review, see, e.g.,~\cite{Buhler:2013zrp}). The Crab nebula and the pulsar inside the nebula (together referred to as ``the Crab'') are the remnants of the supernova observed in 1054, located in the constellation of Taurus at a distance of about $\unit[2]{kpc}$. Especially for gamma-ray astronomy, the Crab is of considerable interest, as it is the brightest persistent source of photons with energies above the TeV range. In fact, the energy spectrum of the Crab as reported by LHAASO continues without a visible cutoff well into the PeV range, with a maximum observed photon energy of $\unit[1.12]{PeV}$~\cite{LHAASO:2021cbz}. The Galactic Center is another key source which we include in this study. At the center of the Milky Way, at a distance of about $\unit[8.2]{kpc}$ in the constellation Sagittarius in the Southern sky, a supermassive black hole is located (for a review, see, e.g.,~\cite{Genzel:2010zy}). Photons from the Galactic Center have been observed up to energies of $\unit[0.114]{PeV}$~\cite{Albert:2024aaa}, without a visible cutoff. 

In addition to these two prominent objects, we include two more sources from the LHAASO catalog~\cite{LHAASO:2023rpg}: 1LHAASO J1908+0615u is taken as a representative of a number of sources with similar spectral parameters, which give comparable extrapolated UHE photon fluxes. Again, the data do not show a cutoff towards higher energies. In~\cite{LHAASO:2023rpg}, 1LHAASO J1908+0615u has been associated with the pulsar PSR J1907+0602, a radio-faint gamma-ray pulsar inside a bright TeV pulsar wind nebula~\cite{Abdo:2010ht} at a distance of about $\unit[2.4]{kpc}$. Finally, 1LHAASO J2031+4052u* has been selected because the fitted power-law spectrum is particularly hard with a spectral index $\Gamma = 2.13$, leading to high extrapolated fluxes in the UHE regime. This makes 1LHAASO J2031+4052u* a very interesting source in the context of the study presented in this paper, despite detailed spectrum data not yet being publicly available. In~\cite{LHAASO:2023rpg}, 1LHAASO J2031+4052u* is associated with the source LHAASO J2032+4102 from the previous LHAASO catalog~\cite{LHAASO:2021gok}. LHAASO J2032+4102 is the source from which the highest-energy photon ($\unit[1.4]{PeV}$) reported in~\cite{LHAASO:2021gok} was observed. This source was in turn associated with the pulsar PSR 2032+4127 in~\cite{LHAASO:2021gok}, a radio and gamma-ray pulsar in a binary system within the Cygnus OB2 region~\cite{Ho:2016mit}, at a distance of about $\unit[1.3]{kpc}$.

\begin{table*}[p]
\centering
\caption{Parameters of the power-law fits following Eq.~(\ref{eq:pl}) for the four sources shown in Fig.~\ref{fig:measurements}. The values have been taken from~\cite{LHAASO:2023rpg,Albert:2024aaa}. For the calculation of the benchmark scenarios in this work, only the central values are used. A brief discussion of the impact of varying the spectral indices within their respective uncertainties can be found in Sec.~\ref{sec:prospects}.}
\vspace*{2mm}
\begin{tabular}{cccc}
\toprule
 Source &  $\Phi_0\ \unit{[eV^{-1}\,km^{-2}\,yr^{-1}]}$ & $E_0\ \unit{[PeV]}$ & $\Gamma$ \\
 \midrule
 Crab & \multirow{2}{*}{$(1.97\,{\pm}\,0.03)\,{\times}\,10^{-10}$} & \multirow{2}{*}{$0.05$} & \multirow{2}{*}{$3.19\,{\pm}\,0.03$} \\
 (1LHAASO J0534+2200u) & & & \\[1mm]
 Galactic Center & \multirow{2}{*}{$(4.73\,{\pm}\,0.95\,\text{(stat.)}\,\substack{{+}0.25\\{-}0.41}\,\text{(sys.}))\,{\times}\,10^{-10}$} & \multirow{2}{*}{$0.026$} & \multirow{2}{*}{$2.88\,{\pm}\,0.15\,\text{(stat.)}\,{-}\,0.1\,\text{(sys.)}$} \\
 (HAWC J1746$-$2856) & & & \\[1 mm]
 1LHAASO J1908+0615u & $(2.16\,{\pm}\,0.05)\,{\times}\,10^{-10}$ & $0.05$ & $2.82\,{\pm}\,0.03$ \\[1mm]
 1LHAASO J2031+4052u* & $(2.52\,{\pm}\,0.63)\,{\times}\,10^{-12}$ & $0.05$ & $2.13\,{\pm}\,0.27$ \\
\bottomrule
\end{tabular}
\label{tab:powerlaws}
\end{table*}

We focus here on the high-energy ends of the spectra. Hence, we do not include measurements at lower energies below the TeV range. The data have been fitted by the respective Collaborations with a power-law function of the form
\begin{equation}
\Phi_\gamma(E) = \Phi_0 \times \left(\frac{E}{E_0}\right)^{-\Gamma},
\label{eq:pl}
\end{equation}
where $\Phi_0$ denotes the differential photon flux at the pivot energy $E_0$ and $\Gamma$ is the spectral index. The parameters of the power-law fits as reported in~\cite{LHAASO:2023rpg,Albert:2024aaa} for the four sources are given in Tab.~\ref{tab:powerlaws}, and the spectra are displayed in Fig.~\ref{fig:measurements}. We note that in some cases, also a log-parabola function or a power-law function with an exponential cutoff could be used to fit the data, particularly when including the low-energy part. However, we keep to the default power-law fits as given in the aforementioned publications, as they provide a good description of the spectra at the high-energy end. This way, we obtain benchmark numbers for the experimental sensitivity required for giant air-shower observatories to test the UHE luminosity of sources like the selected ones.


\section{Extrapolating the Energy Spectra}
\label{sec:spectra}

In the next step, we extrapolate the spectra of the four sources to higher energies. Such an extrapolation is motivated by the spectra continuing without a cutoff into the PeV range (see the preceding section). We note that the extrapolation of the measured spectra is purely phenomenological, based only on observational data. Currently, there is a strong effort from the theory side to model the PeV photon emission of, for example, pulsar wind nebulae~(see, e.g.,~\cite{WilhelmideOna:2022zmp}), to which a number of the LHAASO PeV $\gamma$-sources could be associated to. While such models may also predict a maximum energy or a cutoff for the photons emitted from such sources (or a subset thereof), observational data are needed to verify these predictions. Even if no such photons were to be found and upper limits below the extrapolation are placed, this provides indications of the maximum energies reached in PeV $\gamma$-sources, which can then be used as input for the models. As an example for this approach, we mention a study published by the Pierre Auger Collaboration, where the absence of an excess of photon-like air-shower events from the direction of the Galactic Center was used to constrain the extrapolation of the spectrum measured by H.E.S.S. into the UHE regime~\cite{PierreAuger:2016ppv}.

\begin{figure}[t]
  \centering
  \includegraphics[width=\columnwidth,]{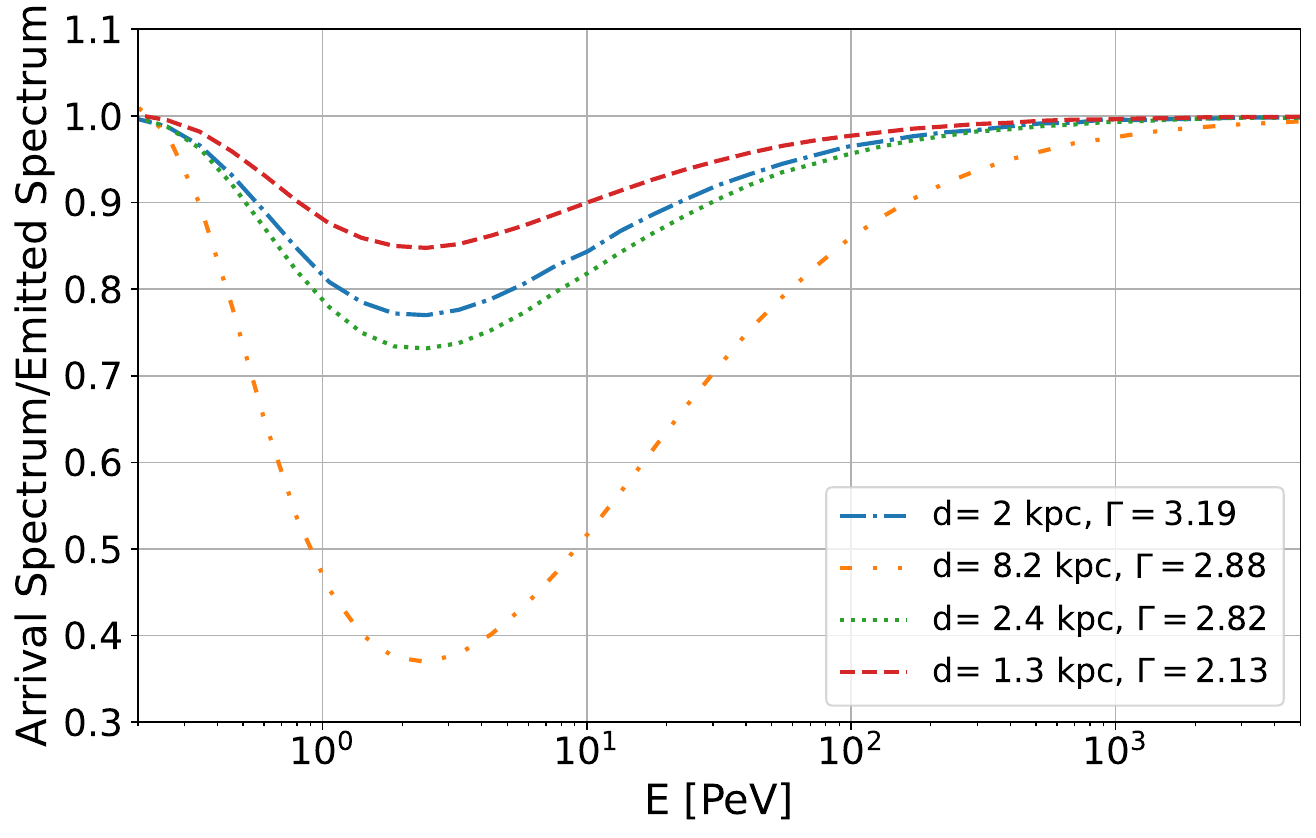}
\caption{Suppression factor due to propagation as a function of the photon energy: The ratio of the spectrum arriving at Earth to the initial spectrum at the source is given for the parameters (distance and spectral index) of the four selected sources.}
\label{fig:propagation} 
\end{figure}

One might wonder whether propagation effects, leaving a power-law spectrum intact up to the maximum energies observed so far, might have a stronger impact in the UHE regime such that the extrapolated flux at Earth would require to be modified. The attenuation length of photons for pair production on the cosmic microwave background (CMB) reaches galactic scales at a few PeV, increasing quickly for smaller and higher energies~\cite{Risse:2007sd}. We checked for the impact of this on the spectra, comparing power-law spectra emitted at the source with the spectra arriving at Earth. In Fig.~\ref{fig:propagation}, based on simulations with CRPropa 3.2~\cite{AlvesBatista:2022vem}\footnote{The simulations were performed using the photon background fields from \cite{Gilmore:2011ks} and \cite{Protheroe:1996si} in addition to the CMB, and with the photon production and electromagnetic cascade propagation according to \cite{Heiter:2017cev}.}, the results are given when adopting the specific distances and spectral indices of each of the four sources. One can see that for the parameters of the three closer sources (distances between $1.3$ and $\unit[2.4]{kpc}$), the effect at UHE is below $20\%$ ($5\%$) at $\unit[10]{PeV}$ ($\unit[100]{PeV}$) and, thus, fairly small. For the distance of the Galactic Center ($\unit[8.2]{kpc}$), the flux is reduced by $50\%$ at $\unit[10]{PeV}$, 30\% at $\unit[30]{PeV}$ and $15\%$ at $\unit[100]{PeV}$. For the approximative benchmark numbers we are interested in here, the overall impact is still comparatively small and propagation effects are therefore not accounted for in the following. They should be kept in mind, however, when focusing on the PeV region or on distant sources. We note that a measurement of such a spectral suppression as expected from the CMB interaction, i.e.\ a dip at a few PeV and a recovery towards the UHE range, would be very interesting in itself, as it is sensitive to fundamental physics (pair production in a highly boosted system) and to astrophysics (e.g., source distance and presence of intervening fields and matter).

Returning to the extrapolation, we are interested in the (integral) number of photons $n_\gamma$ (per unit area and time) reaching Earth from these sources with energies above a given threshold energy $E_\text{thr}$. Integrating Eq.~(\ref{eq:pl}) yields
\begin{equation}
n_\gamma\left(E \geq E_\text{thr}\right) = \frac{\Phi_0\,E_0}{\Gamma-1}\times\left(\frac{E_\text{thr}}{E_0}\right)^{1-\Gamma}.
\label{eq:ngamma}
\end{equation}

\begin{figure*}[t]
  \centering
  \includegraphics[width=0.98\textwidth,]{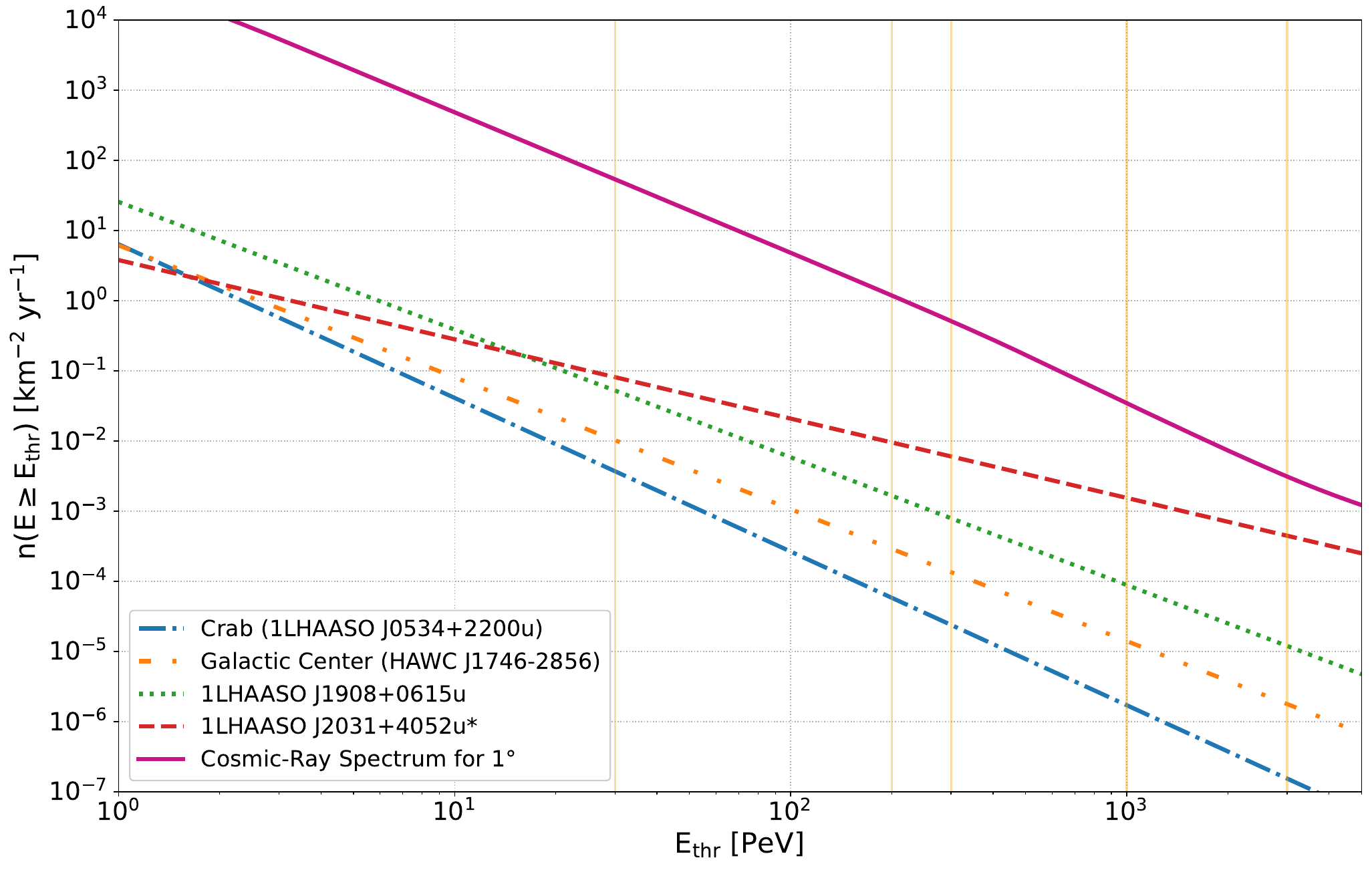}
\caption{(Integral) number of particles (photons and cosmic rays, respectively) reaching Earth per unit area and time with energies above the threshold energy $E_\text{thr}$. $n_\gamma$ follows Eq.~(\ref{eq:ngamma}) with the parameters for the four sources from Tab.~\ref{tab:powerlaws}, $n_\text{CR}$ is given for a circle of $1^\circ$ radius in the sky. The solid vertical (orange) lines indicate the energy thresholds from Tab.~\ref{tab:experiment}.}
\label{fig:extrapolatedspectra} 
\end{figure*}

In Fig.~\ref{fig:extrapolatedspectra}, $n_\gamma$ is shown as a function of the threshold energy for the four sources, following Eq.~(\ref{eq:ngamma}) with the parameters from Tab.~\ref{tab:powerlaws}. Comparing the extrapolations, one finds that the expected number of photons from 1LHAASO J1908+0615u is largest until about $\unit[20]{PeV}$, due to the larger value of $\Phi_0$. At higher energies, 1LHAASO J2031+4052u* takes over due to the smaller value of $\Gamma$. For both sources, the number of photons is $n_\gamma \simeq 0.3 \, \unit{km^{-2}\,yr^{-1}}$ above $\unit[10]{PeV}$, and $0.02$ ($0.006$) $\unit{km^{-2}\,yr^{-1}}$ for 1LHAASO J2031+4052u* (for 1LHAASO J1908+0615u) above  $\unit[100]{PeV}$.
The Crab exhibits the steepest spectrum, which leads to the lowest extrapolated number of UHE photons among the four selected sources. Above $\unit[10]{PeV}$, $n_\gamma$ is about one order of magnitude smaller for the Crab than for 1LHAASO J1908+0615u and 1LHAASO J2031+4052u* (and two orders of magnitude smaller compared to the latter above $\unit[100]{PeV}$). For the Galactic Center, the spectral index is similar to 1LHAASO J1908+0615u, but due to $\Phi_0$ (here at a pivot energy of $\unit[0.026]{PeV}$ instead of $\unit[0.05]{PeV}$ as for the other three sources), $n_\gamma$ is about a factor ${\sim}5$ below 1LHAASO J1908+0615u.

To estimate the prospects for an observation of UHE photons from these sources---or more general, of any potential source exhibiting a similar energy spectrum---, the background from cosmic rays has to be taken into account. The challenge in searches for photons using air showers lies in distinguishing photon-initiated showers from those induced by (charged) cosmic-ray particles such as protons and heavier nuclei. The cosmic-ray flux has been measured in the energy region of interest for example by IceCube/IceTop~\cite{IceCube:2020yct} and, towards the highest energies, by the Pierre Auger Observatory~\cite{PierreAuger:2021hun}. The cosmic-ray spectrum is, over a wide energy range, essentially a broken power-law, with three distinct breaking points at $\unit[3]{PeV}$ (the ``knee''), where the spectral index changes from $2.7$ to $3.0$, at $\unit[500]{PeV}$ (the ``second knee''), where the spectrum further steepens to a spectral index of $3.3$, and at $\unit[4900]{PeV}$ (the ``ankle''), where the spectrum gets flatter again with a spectral index of $2.52$. Towards the highest energies, beyond the energy region we are interested in here, the flux of cosmic rays is heavily suppressed~\cite{PierreAuger:2021hun}, which we also take into account in the calculations for completeness. The (integral) number of cosmic-ray particles $n_\text{CR}$ (per unit area and time) reaching Earth with energies above a given threshold energy $E_\text{thr}$ is also included in Fig.~\ref{fig:extrapolatedspectra}. Since we want to estimate the cosmic-ray background when searching for UHE photons from a point source, $n_\text{CR}$ is given for a circle of $1^\circ$ radius in the sky, reflecting the typical angular resolution of current air-shower experiments. Comparing $n_\text{CR}$ to $n_\gamma$ for the sources investigated here, one finds that the number of cosmic-ray particles can be a few orders of magnitude above the expected number of photons, the impact of which we will briefly discuss in the next section.


\section{Prospects for Observing UHE Photons}
\label{sec:prospects}

\begin{table*}[t]
\centering
\caption{Expected (integral) numbers of photons with energies above the energy threshold for different combinations of detector area and energy threshold, all based on the parameters of detector systems of the Pierre Auger Observatory~\cite{PierreAuger:2015eyc,PierreAuger:2022gkb}, for the four exemplary sources shown in Fig.~\ref{fig:extrapolatedspectra}. The numbers are given for a total measurement time of ten years. Also given are the expected numbers of background cosmic-ray particles per point source from a circle of $1^\circ$ radius for the given detector parameters. The two last rows refer to hybrid measurements, for which a reduced duty cycle of $\unit[15]{\%}$ has been taken into account.}
\vspace*{2mm}
\begin{tabular}{ccccccc}
\toprule
\multicolumn{2}{c}{\multirow{2}{*}{Exemplary detector parameters}} & \multicolumn{4}{c}{$n_\gamma(E \geq E_\text{thr})\times A \times \unit[10]{yr}$} & $n_\text{CR}(E \geq E_\text{thr})$ \\
 \multicolumn{2}{c}{} & \multicolumn{4}{c}{for a source like} & $\times A \times \unit[10]{yr}$\\
\multirow{2}{*}{Area $A\ \unit{[km^2]}$} & Energy threshold & Crab & Galactic & 1LHAASO & 1LHAASO & \multirow{2}{*}{per point source}\\
 & $E_\text{thr}\ \unit{[PeV]}$ & & Center & J1908+0615u & J2031+4052u* & \\
\midrule
$1.95$ & $30$ & \multirow{2}{*}{$0.072$} & \multirow{2}{*}{$0.20$} & \multirow{2}{*}{$1.0$} & \multirow{2}{*}{$1.6$} & \multirow{2}{*}{$1082$} \\
\multicolumn{2}{c}{\footnotesize{(cf. Auger SD $\unit[433]{m}$)}} & & & \\
$27.5$ & $300$ & \multirow{2}{*}{$0.0066$} & \multirow{2}{*}{$0.037$} & \multirow{2}{*}{$0.22$} & \multirow{2}{*}{$1.6$} & \multirow{2}{*}{$146$} \\
\multicolumn{2}{c}{\footnotesize{(cf. Auger SD $\unit[750]{m}$)}} & & & \\
$3000$ & $3000$ & \multirow{2}{*}{$0.0046$} & \multirow{2}{*}{$0.053$} & \multirow{2}{*}{$0.36$} & \multirow{2}{*}{$13.3$} & \multirow{2}{*}{$96$} \\
\multicolumn{2}{c}{\footnotesize{(cf. Auger SD $\unit[1500]{m}$)}} & & & \\
\midrule
$27.5$ & $200$ & \multirow{2}{*}{$0.0024$} & \multirow{2}{*}{$0.012$} & \multirow{2}{*}{$0.068$} & \multirow{2}{*}{$0.39$} & \multirow{2}{*}{$47$} \\
\multicolumn{2}{c}{\footnotesize{(cf. Auger Hybrid, FD + SD $\unit[750]{m}$)}} & & & \\
$3000$ & $1000$ & \multirow{2}{*}{$0.0077$} & \multirow{2}{*}{$0.063$} & \multirow{2}{*}{$0.40$} & \multirow{2}{*}{$6.9$} & \multirow{2}{*}{$148$} \\
\multicolumn{2}{c}{\footnotesize{(cf. Auger Hybrid, FD + SD $\unit[1500]{m}$)}} & & & \\
\bottomrule
\end{tabular}
\label{tab:experiment}
\end{table*}

In the previous section, we calculated $n_\gamma$ in the UHE regime for the four sources. To estimate the potential for an actual observation with air-shower experiments, the parameters of the detector have to be taken into account. For a reasonable estimate, mainly the area covered by the experiment and the energy threshold are relevant. Here, we use the parameters of the Pierre Auger Observatory as a numerical example, as it is the largest air-shower experiment in the world. Naturally, the results can be transferred to any current or future air-shower experiment, for example the Telescope Array experiment~\cite{TelescopeArray:2008toq}, the Global Cosmic-Ray Observatory GCOS~\cite{Horandel:2021prj}---a next-generation giant air-shower observatory intended to succeed current-generation experiments like the Pierre Auger Observatory---, or the PEPS (Probing Extreme PeVatron Sources) project~\cite{Maris:2023anl}, a proposed specialized instrument targeted at PeV $\gamma$-sources, utilizing a dense array of layered water Cherenkov detectors~\cite{Letessier-Selvon:2014sga}. We emphasize that the goal here is not to estimate the prospects of observing, e.g., \textit{the} Crab at ultra-high energies with \textit{the} Pierre Auger Observatory, but rather the prospects of measuring photons from \textit{a source like the} Crab, i.e., one that exhibits a similar energy spectrum, with \textit{a detector like the} Pierre Auger Observatory. The detector could be located anywhere in the world, just as the source could be located anywhere in the sky. For specific sources and detectors, one would have to take into account the visibility of the source from the location of the detector, but we neglect this here, assuming that the exemplary source is always within the field of view of the exemplary detector. As mentioned before, we therefore obtain benchmark scenarios from which giant air-shower experiments can determine the sensitivity they should aim for to either observe UHE photons from PeV $\gamma$-sources or provide significant constraints on their flux.

The expected (integral) numbers of photons with energies above the energy threshold for different exemplary combinations of detector area and energy threshold, all based on the parameters of detector systems of the Pierre Auger Observatory, are given in Tab.~\ref{tab:experiment} for sources like the ones displayed in Fig.~\ref{fig:measurements}. The main surface detector array (SD) of the Pierre Auger Observatory covers $\unit[3000]{km^2}$, with a distance of $\unit[1500]{m}$ between neighbouring detector stations, resulting in an energy threshold of ${\sim}\unit[3000]{PeV}$. Within the main array, there are also smaller sub-arrays, covering $\unit[27.5]{km^2}$ ($\unit[1.95]{km^2}$) with energy thresholds of about $\unit[300]{PeV}$ ($\unit[30]{PeV}$). The lower energy thresholds originate from smaller distances between the detector stations ($\unit[750]{m}$ and $\unit[433]{m}$, respectively). We assume a total measurement time of ten years. In addition, we give the expected numbers of photons for hybrid measurements employing different surface detector arrays in combination with fluorescence detectors (FD). Hybrid measurements have a lower energy threshold compared to SD-only measurements, however, they can only be conducted in clear, moonless nights. This reduces the duty cycle to about $\unit[15]{\%}$, which is taken into account in Tab.~\ref{tab:experiment}.

As expected from the comparison of the four sources in Fig.~\ref{fig:measurements}, the Crab leads to the smallest and 1LHAASO J2031+4052u* to the largest numbers. For three sources, the largest numbers are achieved for the lowest-energy threshold. In case of a hard-spectrum source like 1LHAASO J2031+4052u*, the number increases for the high-energy threshold, as the reduced flux is overcompensated by the gain in detector area. Comparing the hybrid setups with the ground array ones, there is a certain compensation between the reduced threshold energy (equalling more photons) and the reduced duty cycle.

Overall, the extrapolated photon numbers for a single source with a given setup are small. Only for 1LHAASO J2031+4052u*, and for the lowest-energy threshold for 1LHAASO J1908+0615u, values are of $\mathcal{O}(1)$ or above. The largest value is obtained for 1LHAASO J2031+4052u* for the highest energy threshold of $\unit[3000]{PeV}$ (and a corresponding detector area of $\unit[3000]{km^2}$) with an extrapolated number of $13.3$ photons over a measurement time of ten years. With a hybrid measurement, $6.9$ photons would be expected above $\unit[1000]{PeV}$.

These numbers might well be reduced, as mentioned before, when taking the visibility of a specific source for a given detector location into account. As an example, while the Galactic Center culminates close to the zenith for the location of the Pierre Auger Observatory, the Crab reaches only low elevations (at most ${\sim}33^\circ$ above horizon). Also, the effective detector area is reduced with increasing zenith angle of the source.

On the other hand, the numbers could be increased when combining (``stacking'') observations from several sources. As mentioned before, there are several LHAASO sources with similar spectral parameters as 1LHAASO J1908+0615u. Partly, also the numbers related to different detector setups could be added: for instance, the data sets for the $\unit[433]{m}$ and $\unit[750]{m}$ SD arrays are largely independent from each other due to the different energy thresholds.

The photon numbers in Tab.~\ref{tab:experiment} are based on the central values of the observations as given in Tab.~\ref{tab:powerlaws}. The relative flux uncertainties are $\Delta\Phi_0/\Phi_0 < 30\,\%$ and, thus, comparatively small. This also holds for the uncertainties of the spectral index of $\Delta\Gamma = 0.03$ for the Crab and 1LHAASO J1908+0615u. However, for 1LHAASO J2031+4052u* ($\Delta\Gamma = 0.27)$ and the Galactic Center ($\Delta\Gamma = 0.25$, adding up statistical and systematic uncertainties), varying the spectral index within uncertainties can alter the extrapolated photon numbers by, e.g., about an order of magnitude for a threshold of ${\sim}\unit[300]{PeV}$ (smaller/larger number of photons for softer/harder spectrum). The reduced photon numbers for 1LHAASO J2031+4052u* (and the increased photon numbers for the Galactic Center) would then be comparable to those of 1LHAASO J1908\-+0615u. The extreme case of increased photon numbers for 1LHAASO J2031+4052u* for a very hard spectrum could actually be in reach for current giant-air shower observatories.

Also listed in Tab.~\ref{tab:experiment} are the expected numbers of (background) cosmic-ray particles (per point source), calculated from the energy spectrum of cosmic rays (see Sec.~\ref{sec:spectra}) for the same combinations of detector area and energy threshold and the same measurement time of ten years. These numbers can be compared to the expected numbers of photons. Assuming similar detection efficiencies for photons and cosmic-ray particles, a background suppression to a level of about $10^{-3}$ to $10^{-4}$ is needed to make a discrimination between photons and cosmic-ray particles feasible. Current photon searches at ultra-high energies (and below) already achieve background suppressions to a level of $10^{-4}$ and better~\cite{PierreAuger:2022gkb}. We mention here as an example a search for photons with energies above $\unit[10^{19}]{eV}$ published by the Pierre Auger Collaboration, using only data from the surface detector array of water Cherenkov detectors~\cite{PierreAuger:2022aty}. The analysis was applied to a search sample of about $48{,}000$ air-shower events, yielding $16$ candidate events. Assuming these are all background events would lead to an overall background suppression to the level of $3\,{\times}\,10^{-4}$. Naturally, with more complex detectors, a better background suppression can be reached. For example, using additional information from dedicated muon detectors can improve the background suppression to the level of $10^{-5}$~\cite{Gonzalez:2020ztr}.


\section{Summary and Outlook}
\label{sec:summary}

Motivated by the recent observations of PeV photons by LHAASO and HAWC, we studied if (and under what conditions) giant air-shower observatories could contribute to testing the UHE luminosity of PeV $\gamma$-sources. Selecting four exemplary sources, benchmark numbers for their photon fluxes were obtained from extrapolating the power-law fits to the UHE regime (above $\unit[10]{PeV}$). Possible propagation effects were investigated and found to be reasonably small for the purpose of this work. Adopting the present parameters of the Pierre Auger Observatory as an example, the extrapolated numbers of photons were calculated for the different detector systems employed. In the spirit of providing benchmark estimations, we did not check the prospects of observing a specific source with the Pierre Auger Observatory, but the prospects of testing the UHE luminosity from sources with energy spectra similar to the ones selected with a detector configuration like the Pierre Auger Observatory.

The extrapolated numbers of photons are small. The values are of $\mathcal{O}(1)$ or above, adopting a measurement time of ten years, only for a source like 1LHAASO J2031+4052u* due to its hard spectrum, and for a source like 1LHAASO J1908+0615u at the presently lowest-energy threshold of $\unit[30]{PeV}$. These numbers will decrease when accounting for realistic visibility conditions. They can be increased when source stacking is applied---for instance, several sources like 1LHAASO J1908+0615u are observed already. Overall, the test of the UHE luminosity of the detected PeV $\gamma$-sources by giant air-shower observatories appears to be a challenge at present.

To increase the sensitivity, apart from a larger detector aera, lowering the energy threshold would be beneficial. As an example, let us assume a detector area of $\unit[10]{km^2}$ and a threshold energy of $\unit[10]{PeV}$. Then, the numbers (for ten years measurement time) would be about $38.5$ (1LHAASO J1908+0615u), $28.0$ (1LHAASO J2031+4052u*), $8.1$ (Galactic Center), and $4.1$ (Crab), reaching a statistically more realistic level. The corresponding cosmic-ray background would amount to ${\sim}48300$, requiring a background suppression at the level of $10^{-3}$, which seems feasible.

LHAASO and HAWC, located in the Northern hemisphere, continue to observe PeV $\gamma$-sources and, probably, detect further ones. A complementary detector in the Southern hemisphere, like the planned SWGO~\cite{Hinton:2021rvp}, will likely lead to the discovery of more such sources. Testing their UHE luminosity is potentially very rewarding but turns out to be rather challenging with present setups. Specific detector concepts, like the one considered in~\cite{Maris:2023anl}, appear to be necessary to further explore the UHE frontier of these exciting inhabitants of our Galaxy.


\section*{Acknowledgements}

This work was supported by the German Research Foundation (DFG Project No. 508269468). We thank our colleagues from the Pierre Auger Collaboration for fruitful discussions.


\bibliographystyle{elsarticle-num}
\bibliography{bibliography}

\end{document}